# Free-space optical neural network based on thermal atomic nonlinearity

ALBERT RYOU,[1,*] JAMES WHITEHEAD,[1] MAKSYM ZHELYEZNYAKOV,[1] PAUL ANDERSON,[2,3] CEM KESKIN,[4] MICHAL BAJCSY[3,5] AND ARKA MAJUMDAR[1,6]

[1]*Department of Electrical and Computer Engineering, University of Washington, Seattle, Washington 98195, USA*
[2]*Department of Physics and Astronomy, University of Waterloo, Waterloo, Ontario, Canada*
[3]*Institute of Quantum Computing, University of Waterloo, Waterloo, Ontario, Canada*
[4]*Google, Mountain View, California 94043, USA*
[5]*Department of Electrical and Computer Engineering, University of Waterloo, Waterloo, Ontario, Canada*
[6]*Department of Physics, University of Washington, Seattle, Washington 98195, USA*
*\*albertryou@gmail.com*

**Abstract:** As artificial neural networks (ANNs) continue to make strides in wide-ranging and diverse fields of technology, the search for more efficient hardware implementations beyond conventional electronics is gaining traction. In particular, optical implementations potentially offer extraordinary gains in terms of speed and reduced energy consumption due to intrinsic parallelism of free-space optics. At the same time, a physical nonlinearity – a crucial ingredient of an ANN – is not easy to realize in free-space optics, which restricts the potential of this platform. This problem is further exacerbated by the need to perform the nonlinear activation also in parallel for each data point to preserve the benefit of linear free-space optics. Here, we present a free-space optical ANN with diffraction-based linear weight summation and nonlinear activation enabled by the saturable absorption of thermal atoms. We demonstrate, via both simulation and experiment, image classification of handwritten digits using only a single layer and observed 6% improvement in classification accuracy due to the optical nonlinearity compared to a linear model. Our platform preserves the massive parallelism of free-space optics even with physical nonlinearity, and thus opens the way for novel designs and wider deployment of optical ANNs.

## 1. Introduction

Artificial neural networks (ANNs) have recently proven phenomenally successful in tasks such as image, sound, and language recognition and translation [1]. The increasing deployment of ANNs, from facial recognition on smartphones to self-driving cars, has brought new attention to improving their hardware implementation in terms of speed, energy consumption, and latency [2]. In contrast to conventional electronics-based platforms, optical implementations stand out due to light's intrinsically massive parallelism. For instance, the ability of a simple lens to carry out a two-dimensional Fourier transform with zero energy has long been utilized in optical signal processing [3]. Especially, free-space optics (FSO) with an aperture area $A$ and wavelength $\lambda$ can potentially provide an extremely large number of information channels $\sim A/\lambda^2$, thanks to the availability of two spatial dimensions.

One of the biggest hurdles for an optical implementation of an ANN, however, is the lack of physical optical nonlinearity. While the parallelism of FSO naturally lends itself to carrying out linear operations, the lack of corresponding parallel nonlinearity without requiring high-powered lasers or active optical components has led to a multitude of non-FSO workaround solutions. Shen *et al.* demonstrated an electronic-optical hybrid neural network, in which the output of an integrated photonic mesh was outsourced to an external computer for nonlinear processing [4]. Nevertheless, it was shown by Colburn *et al.* that the benefits of such a design with repeated data conversions between the optical and the electronic domains were severely limited due to large power consumption and latency incurred during signal transduction [5]. Furthermore, an integrated photonics platform foregoes the intrinsic parallelism of two-dimensional FSO. For example, Feldmann *et al.* demonstrated a fully optical spiking network with on-chip phase-change materials; but scaling up the number of neurons beyond a few waveguides remains technically challenging [6]. While wavelength division multiplexing (WDM) has been theoretically proposed as a promising route to mitigating the challenges for scaling the number of waveguides [7], such methods need to stabilize high-Q ring resonators under thermal fluctuations, leading to excess energy consumption. Moreover, a large number of additional control operations are needed to serialize the 2D image data stream and multiplex those data to encode in wavelengths, all of which will need an excess amount of energy. Another recent promising research direction is to completely avoid nonlinearity and employ multiple diffractive layers for classification [8,9]. While such an approach provided impressive classification results for MNIST dataset for a linear network combined with logistic regression, the lack of nonlinearity poses a serious question about its generalizability to solving more complicated tasks.

Recently, Zuo *et al.* presented an FSO neural network, where the nonlinearity comes from the electromagnetically induced transparency in ultracold atoms [10]. Besides extensive laboratory setup for trapping and cooling atoms, the need to hand off the data from one laser to another prevents the extension of this method to having multiple hidden layers. In a similar vein, the quantum well exciton-polariton based nonlinear activation requires a cryostat and is difficult to scale [11].

In this paper, we propose and demonstrate a fully optical ANN that utilizes the optical nonlinearity from thermal atomic vapor. Specifically, we exploit the saturable absorption behavior of room-temperature rubidium atoms housed in a vapor cell. We observed the nonlinearity in a single pass without any cavity, which allows point-by-point nonlinear activation of an incident image [12]. For the linear operations, we employ the diffractive model, where phase masks directly set the trainable weights of the neural network [8]. We emphasize that both the linear and the nonlinear components of our neural network operate on a "pixel-by-pixel" basis, within the diffraction-induced limit set by the propagation length, thus preserving and fully exploiting the intrinsic massive parallelism of FSO. Via numerical simulations, we observed an increase in classification accuracy in a single linear layer ANN by 10% due to the atomic nonlinearity. Following the training of our optical neural network in simulation using experimentally relevant parameters, we experimentally demonstrate an image recognition task of handwritten digits using a spatial light modulator (SLM). We observed an increase in classification accuracy by 6% in experiment with the addition of the nonlinear layer. We attribute the moderate classification accuracy (~33%) of our experimental system to using only a single diffractive linear operation, currently limited by the number of SLMs in our setup. Our work combining machine learning with optics and atomic physics opens a new front in the on-going effort to advance optical ANN theory and hardware.

## 2. Optical neural network architecture

### 2.1 Overview

A typical deep neural network consists of multiple layers of neurons. Except for those in the first layer, each neuron receives input signals from neurons in the previous layer. Excluding batch normalization, the neuron takes the sum of the signals multiplied by adjustable weights and performs a nonlinear operation, the output of which subsequently becomes an input signal for one or more neurons in the following layer.

Many variations on the neural network architectures exist, along with different training algorithms, for specific applications. For a typical image classification task under supervised learning, the network is presented with a set of training data and corresponding labels. By repeatedly comparing the result of the output against the labels, the network can gradually adjust its weights until finally they converge on an optimum solution.

Our optical neural network follows a similar architecture: a two-dimensional, monochromatic wavefront containing the input data propagates sequentially through a series of linear and nonlinear layers before being imaged on a camera. However, as explained earlier due to limited number of available SLMs, we only implemented one single layer that combines the input and one layer of neurons. Below, we describe each component and its physical implementation.

### 2.2 Input layer

The input layer is the direct representation of two-dimensional data encoded as spatially varying intensity of light, or an image. In order to convert electronic data into optical images, we use an SLM, which can manipulate the amplitude, phase, or both of an incident laser beam's wavefront. The use of coherent, monochromatic light is crucial for the reported optical network, since we utilize diffraction and light-matter interaction to perform both linear and nonlinear operations, as will be described next.

### 2.3 Linear layer

In a generic ANN, the role of a linear layer is to perform summation of signals from a previous layer with adjustable weights before passing it off to a nonlinear layer. A direct implementation of matrix-matrix multiplications in free-space optics is possible but complex and requires many optical elements [3]. Instead, we adopt an alternative approach, in which the linear layer is implemented by first elementwise multiplying an image with a phase mask and then letting the image propagate in free space. The first step is enabled by the SLM, which can directly display the product of an input image with the phase mask. The second step allows the signals of neighboring pixels of the image to mix due to diffraction. Such a diffractive model was demonstrated for several phase masks in the THz regime [8]. The amount of mixing depends on the propagation distance, the wavelength of the image, and the spatial frequency spectrum of the image. We note that it is difficult to map such a phase-mask-based approach to a traditional convolutional or fully connected layers used in ANN. However, as the pixels mix with the neighboring pixels, the operation is effectively a convolution operation, the kernel of which depends on the propagation length. However, using a stack of diffractive optics [8], metasurfaces [13], or even more than one SLM, multiple layers can be implemented.

### 2.4 Nonlinear layer

The nonlinear layer is implemented by an evacuated vapor cell filled with rubidium atoms. The phenomenon of saturable absorption is briefly outlined here and further detailed in Appendix. When a near-resonant photon is incident on an atom, the atom absorbs the photon and reaches an excited state. After a brief time that is inversely proportional to the atomic linewidth, the excited atom emits a photon and returns to the ground state. The emitted photon travels in a random direction and is "lost" from the undisturbed wavefront that continues to propagate in the original direction. Thus, for a fixed density of atoms, a low-intensity beam passing through the gas becomes attenuated. On the other hand, a high-intensity beam can excite all the available atoms, saturating the medium. The input-output curve of an optical beam of varying intensity thus exhibits a nonlinear shape (see Fig. 4), similar to the nonlinear activation function type "SmoothReLU" commonly used in machine learning. The key to our nonlinear layer is the fact the saturation of atoms is a local effect, and thus, different parts of an incident image, which can be viewed as a collection of multiple beams with each beam denoting one pixel, undergo the nonlinear activation independently.

### 2.5 Output layer

The optical signal after the vapor cell is imaged on a CCD camera. The intensity pattern of the captured image becomes a direct representation of the final output of the neural network. For an image classification task with multiple categories, we can pre-define certain physical locations on the camera plane to correspond to those categories. These locations then can be read by either a human or a computer to identify the categories.

We note that the absolute squaring operator inherent in taking the intensity is in itself a nonlinear process; however, as it is bound to the final measurement, we take it as part of the output layer and only refer to the independent saturable absorption layer as our nonlinearity.

## 3. Simulation of a two-layer optical neural network for image classification

While atomic vapors provide a nonlinear input-output relationship, it is not clear a priori whether such a nonlinear function will be useful for an optical neural network, especially given there is no energy gain in the system, only loss. To probe the efficacy of the saturable absorption nonlinearity in thermal atoms, we first simulate a two-layer optical neural network: one linear layer (to be implemented by an SLM) and one nonlinear layer (to be implemented by the saturable absorption nonlinearity). We focused on the classic image classification of handwritten digits from the MNIST database. The goal is to define an optical model, train it entirely offline, and implement the trained neural network as closely as possible in experiment. In this section, we describe the training procedure, including the use of the atomic nonlinearity, and discuss the simulation results.

The raw input data are 8-bit, $28 \times 28$ pixels images of handwritten digits. Before feeding them to the model, we make the following modifications. First, in order that the image remains reasonably collimated during tens-of-centimeters-long free-space propagation in the experiment, we rescale the dimensions from the original $28 \times 28$ pixels to $300 \times 300$ pixels. Second, we further embed the $300 \times 300$ pixel image within a larger $600 \times 600$ pixels image, with the area outside the image having zero value. This larger dimension allows us to directly employ the angular spectrum method without applying any bandlimit [13]. Finally, all the values of the image-pixels are normalized so that the maximum pixel value is one. The size of the pixel is set to $8\ \mu m$ to match the physical pitch of our SLM. The first operation on the modified input is elementwise multiplication by a phase mask, which consists of an array of complex numbers whose magnitude is unity and whose phase $\phi(x, y)$ is a trainable variable.

After the phase mask, the image is propagated along the optical axis by a distance $z_o$, which is a hyperparameter for our neural network, via the angular propagation method. The angular propagation method consists of decomposing a given wavefront into plane-waves traveling in different directions, applying a $z_o$-dependent transfer function to each plane-wave, and finally reconstructing the new wave. Computationally, the process involves a pair of forward and inverse fast Fourier transforms along with a Hadamard product with a matrix in between the pair [12].

After propagation, the image undergoes a nonlinear activation. The nonlinearity is a function of the optical intensity, so we take the absolute square of the image field, apply a nonlinear function, and take the square root, all while preserving the phase of the original wavefront. The functional form of the nonlinearity is derived in Appendix; the nonlinear parameters were determined by a calibration process described in Section 4.2.

Finally, for detection, the intensity of the output of the nonlinear layer is elementwise multiplied by a special detector layer that defines where the light of a given MNIST digit should go. In our simulation, the detector layer consists of ten circles that are equidistant from the center. The result of the matrix multiplication is a list of ten numbers, each of which is the sum of the image intensity values within the circle. The maximum number, indicating the location with the highest intensity, is the final output of the neural network for the given sample image.

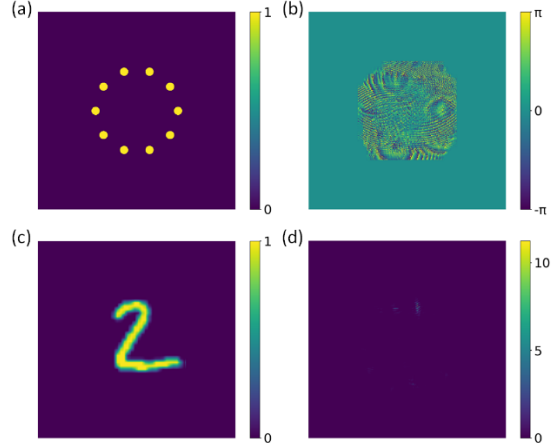

Fig. 1. Trained ONN. (a) The detector layer determines the location, where the light from the individual digits should be focused. The layout of the layer is a hyperparameter in our training. Here, each label corresponds to one bright circle (radius = 100 μm) located 1 mm from the center of the image. The "0" label is on the positive x-axis, and the rest of the labels are located sequentially counterclockwise on a circle. (b) Trained phase mask. (c) Sample input image. (d) Output of the neural network for the sample input shown in (c). For training, the neural network calculates the intensity at each label location and returns the highest-intensity label as its prediction. All images have dimensions of 600 × 600 pixels, which correspond to 4.8 × 4.8 mm.

The entire model was defined and trained using TensorFlow 2.0 on AWS EC2. For training, we used 10,000 training images, 1000 test images, and 50 epochs, and chose Adam for optimization. The images are taken from the MNIST database. Figure 1 shows the results of a trained neural network. Figures 1a and 1b show the detector layer and the trained phase mask. The general layout of the light locations is a hyperparameter. Figures 1c and 1d show the sample input and the output of the neural network for the input. As can be seen, the location corresponding to the "2" label has the highest intensity.

Figure 2 shows the accuracy of the neural network versus the number of training epochs. In order to test the efficacy of our nonlinearity, we simulate a network without any nonlinearity (linear model) as well as the one containing the nonlinear layer (nonlinear model). As can be seen, in both cases, the accuracy rises quickly during the first few epochs and reaches a steady state. After 50 epochs, the accuracy is 74.4% for the linear model and 84.4% for the nonlinear model, showing a significant improvement. We emphasize that these accuracy values are significantly lower than the state-of-the-art ANNs, only because we have only one linear layer. Using multiple linear layers, our model can reach ~100% accuracy for MNIST digits. However, multiple layers will be infeasible to experimentally probe in our lab because of the limited resource. Nevertheless, the increase in the classification accuracy due to the thermal atomic nonlinearity clearly shows that the physical optical nonlinearity is suitable for implementing an ANN.

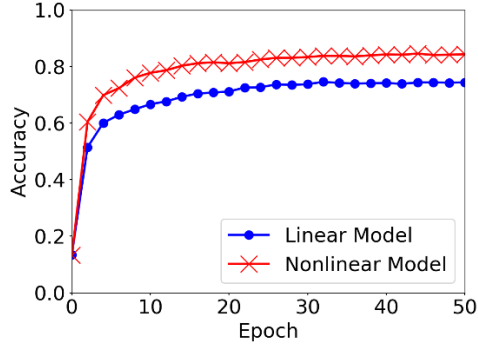

Fig. 2. Accuracy vs epoch for the linear model (blue dot) and the nonlinear model (red x).

## 4. Experimental results

*4.1 Setup*

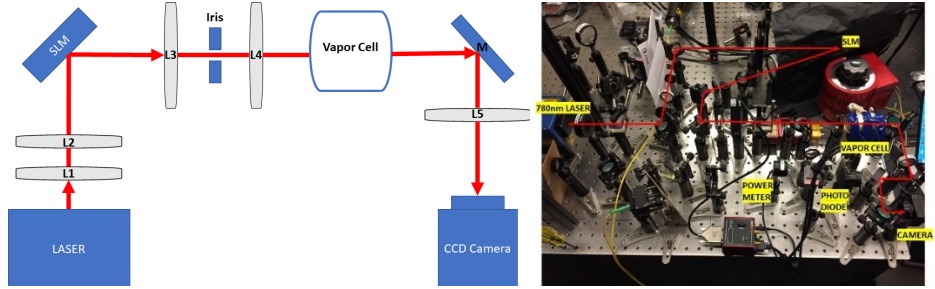

Fig. 3. Experimental setup. (a) Cartoon layout of the setup. The focal lengths of the lenses are: L1 = 50 mm; L2 = 300 mm; L3 = 150 mm; L4 = 150 mm; L5 = 100 mm. M indicates a flat mirror. (b) Photograph of the experiment.

Figure 3 shows the layout of our experimental setup in addition to a photograph. The source of the monochromatic light is a 780-nm-wavelength laser (Toptica DL Pro), whose wavelength is fine-tuned to be resonant with the $5S_{1/2} \rightarrow 5P_{3/2}$ transition of $F_g = 3$ $^{85}$Rb atoms. We first expand the collimated beam with a 1:6 telescope in order to illuminate the SLM (Holoeye Pluto), the size of which is roughly 15.4 mm by 8.6 mm. A pair of 150-mm-focal length lenses then form a relay, between which an iris is placed to pick out the 1st-order diffracted beam. The resulting light, which now encodes the product of a handwritten digit and the trained phase mask, propagates for 100 mm before arriving at the front surface of a 72-mm-long vapor cell (Thorlabs GC25075-RB). The vapor cell can be lowered or raised to easily add or remove the nonlinear layer from the neural network. Furthermore, the vapor cell is wrapped in a variac-controlled heater tape (Omega) for tuning the vapor density via temperature control. During the experiment, the temperature of the cell is maintained at 50°C. Finally, a single 100-mm-focal length lens is used to image the front surface plane of the vapor cell on to a CCD camera (FLIR USB2).

*4.2 Nonlinearity*

Here we describe the calibration process used to derive the nonlinear parameters for both simulation and experiment. The nonlinear input-output curve (see Appendix) can be given by

$$I_{out} = I_{in} \exp\left[-N_{sat}/(1 + I_{in}/I_{sat})\right]$$

where $I_{in}$ and $I_{out}$ are input and output intensities, $N_{sat}$ is the generalized atom density, and $I_{sat}$ is the generalized saturation intensity. To determine the last two parameters, we varied the laser intensity with a waveplate and a polarizer and measured the output with and without the vapor cell. Because we are imaging the entrance plane of the vapor cell, the output measured without the vapor cell can be taken as the input into the vapor cell. From the curve fit, $N_{sat}$ and $I_{sat}$ were determined to be 2.6 and 0.6 µW, respectively, which were then used for simulation in Section 3.

For experiment, it is very difficult to exactly implement the simulated model of the neural network directly due to the attenuation by many optical elements as well as the fact that the vapor cell itself has a finite length on the order of many centimeters. The latter presents a serious challenge, since the simulation assumed that the nonlinear effect took place in a single plane, whereas in the experiment, the nonlinearity occurs over a continuous distance such that a propagating image would be a continuously changing attenuation.

A solution can be found if the intensity is measured not in terms of watts but the pixel value of the CCD camera itself. The parameters $N_{sat}$ and $I_{sat}$ then no longer refer to physical quantities but act as general fit parameters for the nonlinear input-output curve. Figure 4 shows the plot of the average pixel value without (x-axis) and with (y-axis) the vapor cell in place for a sample image with varying overall intensity. The intensity range was chosen to avoid saturation of the camera without having to add an optical attenuator. Once the values of the fit parameters, 1.3 and 520, respectively, were determined, they were used to train the neural network for our experimental results with higher intensity. We emphasize that for actual experiment, we employ the whole dynamic range of the SLM, i.e., pixel values ranging from 0 to 255.

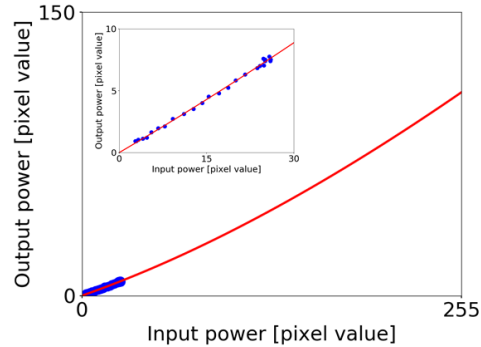

Fig. 4. Nonlinear function showing the input-output curve for the incident intensity. The x-axis is proportional to the input power, or the average pixel valve on the CCD camera without the vapor cell. The y-axis is proportional to the output power, or the average pixel value on the CCD camera with the vapor cell in place. Inset: Zoom-in plot showing the curve fit.

*4.3 Results*

As described before, we trained a new neural network with the nonlinear parameters that were derived directly from the camera, using 10,000 training images and 1000 test images, 100 per digit, which necessitated the adjustment of the input intensity in terms of pixel values rather than milliwatts. The resulting simulation with the experimental parameters yielded a similar-looking phase mask to that of the simulation with the ideal parameters shown in Section 3. However, the predicted accuracy dropped to 66.4% and 66.6% for the linear and the nonlinear networks, respectively, and thus, there was virtually no difference between the two networks in terms of accuracy. While it is possible in theory to achieve the original simulation regime by calibrating each optical element and reconciling simulation and experiment with more advanced techniques such as split-step nonlinear angular propagation [15], the required experimental effort and computational resources would be too great, and so we decided to proceed with experiment.

In our experiment, we used as input the same 1000 test images that were modified as outlined in Section 3. Because our SLM is a phase-only modulator, it cannot directly display an intensity-varying image or a complex field that is the product of the image with a phase mask; hence, we resorted to holography, which allows us to make the complex field in a conjugate plane [16] using only phase control. For detection, we calibrated the CCD camera for image magnification and rotation with separate calibration images. First, we tested the neural network that contained no phase mask. The overall accuracy was 14.7% for the linear network and 14.2% for the nonlinear network. As expected, without the phase mask, there is no significant difference between the two networks, and the accuracy is offset by a small bias near 10% (the baseline accuracy of random prediction).

Next, we repeated the test, this time incorporating the phase mask via the SLM in the neural network. The overall accuracy was measured to be 26.7% for the linear network and 33.0% for the nonlinear network. We attribute the overall reduction in accuracy compared to the simulation results to the imperfect experimental system, including fixing the length between optics, phase error in SLMs, and the finite length of the vapor-cell. However, it is surprising that the accuracy is greater with incorporation of the nonlinearity, whereas the simulation shows similar performance with and without the nonlinearity. We attribute this to the robustness of the nonlinear network to the experimental noise. There is a large body of ongoing research in the machine learning community on the effect of noise in training deep neural network [17-19], and the exact nature of the robustness of our nonlinear optical neural network remains to be investigated. Table 1 summarizes all the accuracy results for both simulation and experiment.

We note that the simulated and experimentally measured efficiencies are significantly lower than the state-of-the-art neural network. However, we have only one layer in our neural network, and we expect the accuracy to increase with a larger number of layers. Currently, our experiment is limited by available resources, e.g., a single SLM, which, while commercially available, is a significant laboratory expenditure. On the other hand, we note that creating multiple layers has several technical challenges of its own, including optical loss in each layer. The reported optical nonlinearity can be tunable by changing the temperature of the atoms, and thus can be tuned for each layer. Moreover, as we are using thermal atoms and we do not rely on cold atoms, the setup is significantly simple. However, optical regeneration techniques will be needed if the depth of the network is too large [20]. Finally, an electronic back end can be used with the optical frontend to enhance the classification accuracy. We emphasize that such an electronic backend requires only one-time transduction, and does not add to the overall latency, as is needed for repeated signal transduction.

**Table 1. Summary of ONN Accuracy in Percentage**

|  | Linear Network | Nonlinear Network |
| --- | --- | --- |
| Simulation with ideal parameters | 74.2 | 84.2 |
| Simulation with experimental parameters | 66.4 | 66.6 |
| Experiment without phase mask | 14.7 | 14.2 |
| Experiment with phase mask | 26.7 | 33.0 |

*4.4 Speed Performance*

Our reported optical ANN uses a commercial liquid crystal-based SLM with 1 million pixels, each pixel with 8-bit precision. The refresh rate of the SLM is ~100Hz, making the effective supported bitrate in the optical ANN as 800Mbps. However, using a grating light valve type of mechanical SLM, we can increase the data rate to ~1Tbps [21]. At that speed, however, we need to ensure a faster detector, e.g., an event-based camera to accommodate ~$\mu sec$ level detector response time [22]. Power consumption of the reported optical ANN primarily comes from the SLM, which is on the order of ~10W. However, as we implement only the inference, we can use a fixed diffractive phase-mask, reducing that energy to zero. For using thermal atom-based nonlinearity, we do not spend any extra energy on either active preparation or maintenance of the nonlinearity. Additionally, the reported optical ANN exploits the full potential of the parallelism offered by FSO, and thus does not require any excess energy needed for time/ wavelength multiplexing. To actuate the nonlinearity, we need a light intensity of ~$16 \mu W/mm^2$, and the required optical power will depend on the SLM pixel size and the optics used to guide the light through the nonlinear thermal atomic vapor. We estimate the average pixel size inside the atomic vapor to be ~$100 \mu m \times 100 \mu m$, making the total required optical power for a million pixel ~$160 mW$. By reducing the channel area to diffraction limited spot (~$1 \mu m \times 1 \mu m$), this power can be reduced to ~$16 \mu W$.

## 5. Conclusion

We have shown that an atomic vapor cell can perform a local nonlinear activation in two-dimensions, and consequently, a fully optical artificial neural network can be implemented for image recognition of handwritten digits. Such a network can handle a large amount of data in parallel. Furthermore, except for the input and the output that are fed and detected by the SLM and the CCD camera, respectively, all data processing occurs in the time light takes the traverse the physical distance of the network. Although the model accuracy of 33% is rather low, our proof-of-concept demonstrates the feasibility of using a simple, off-the-shelf atomic vapor cell as the source of fully parallel optical nonlinearity. Along with another commercially available device, the SLM, the vapor cell solves the enduring challenge of the missing optical nonlinearity that fully exploits the intrinsic massive parallelism of free-space light in two dimensions. Our work is a first step towards creating an all-optical neural network that can handle a massive amount of data and surpass the performance of an electronic neural network.

**Appendix**

**1. Saturable Absorption**

Saturable absorption is a general phenomenon that appears in many different physical systems with discrete energy levels with finite lifetimes. Here we consider a simple system of two-level atoms; a detailed derivation can be found in several resources [23].

Consider a beam of photons passing through a medium with $N$ atoms per unit volume. If the thickness of the medium is $\Delta z$, then the number of atoms per unit area is given by $N \Delta z$. If we now assign an absorption cross-section $\sigma$ to each atom, then $N \sigma \Delta z$ is the fraction of the target area covered by the atoms. It is also the probability that an incident photon will be absorbed by the atoms, or, in the case of many photons, the total fraction of photons that are absorbed. The change in the beam intensity is then $\Delta I/I = -N\sigma\Delta z$, which upon integration, yields Beer's law: $I(z) = I_0 e^{-\kappa z}$, where the absorption coefficient $\kappa = N\sigma$.

If we assume that the atoms have two levels, the ground state and the excited state, then the absorption is given by $\kappa = (N_g - N_e)\sigma$. Imposing the conservation of atom number ($N_{total} =$

$N_g + N_e$) and the conservation of energy (($N_g - N_e)\sigma I = N_e A \hbar \omega$) where the spontaneous decay rate $A = 1/\tau$, we arrive at the steady-state population difference: $N_g - N_e = N_{total}/(1 + I/I_s)$, where we have defined the saturation intensity $I_{sat} = \hbar \omega A/(2\sigma)$.

Thus, the output intensity as a function of the input intensity is given by: $I_{out} = I_{in} \exp[-N_{total}\sigma L/(1 + I_{in}/I_{sat})]$, where $L$ is the effective interaction length of the vapor cell. We use $N_{sat} = N_{total}\sigma L$ and $I_{sat}$ as our nonlinear parameters. We note that for the atomic vapor system, the variable $N_{total}$ can be controlled by changing the temperature of the cell. Thus, the demonstrated nonlinearity is tunable, which can be exploited for multi-layer optical neural network, where the $N_{sat}$ value will be gradually decreased to accommodate the signal loss in each layer.

**Funding.** Washington Research Foundation, Industry Canada, UW Reality Lab, Facebook, Google, Amazon, and Futurewei.

**Acknowledgments.** A. R. acknowledges support from the IC Fellowship. M.Z. acknowledges support from the NSF Graduate Research Fellowship. P. A. acknowledges support from Canada First Excellence Research Fund.

**Disclosures.** The authors declare no conflicts of interest.